\documentclass[a4paper, twocolumn]{article}
\usepackage[utf8]{inputenc}

\usepackage[hmarginratio=1:1,top=32mm,columnsep=20pt]{geometry}

\usepackage{amsmath}
\usepackage{amsfonts}
\usepackage{amssymb}
\usepackage{amsthm}
\usepackage{xcolor}
\usepackage{bm}
\usepackage{mathtools}
\usepackage{graphicx}
\usepackage{subcaption}
\usepackage{float}
\usepackage{commath}
\usepackage[noabbrev]{cleveref}
\usepackage{multirow}
\usepackage{multicol}
\usepackage{caption}
\usepackage{algorithm}
\usepackage{algpseudocode}
\usepackage{url}

\usepackage[
	backend=biber,
	style=numeric,
        maxbibnames=99,
        maxcitenames=1,
        uniquelist=false
	]{biblatex}
\renewbibmacro{in:}{%
  \ifentrytype{article}{}{\printtext{\bibstring{in}\intitlepunct}}}
\DeclareNameAlias{author}{last-first}
\theoremstyle{definition}

\theoremstyle{remark}

\newcommand{\R}{\mathbb{R}}

\usepackage{authblk}
\title{Using Topology to Estimate Structural Similarities of Proteins}
\author[1, 2]{Jørgen Ellegaard Andersen}
\author[3]{Jens Ledet Jensen}
\author[2]{Yuki Koyanagi}
\author[4, 5]{Jakob Toudahl Nielsen}
\author[6]{Rasmus Villemoes}
\affil[1]{Danish Institute for Advanced Study, University of Southern Denmark}
\affil[2]{Centre for Quantum Mathematics, Department of Mathematics and Computer Science, University of Southern Denmark}
\affil[3]{Department of Mathematics, Aarhus University}
\affil[4]{Interdisciplinary Nanoscience Center (iNANO), Aarhus University}
\affil[5]{Department of Chemistry, Aarhus University}
\affil[6]{Prevas A/S}
\date{}
\begin{document}

\twocolumn[
\begin{@twocolumnfalse}
\maketitle  
\begin{abstract}
  \noindent
  An effective model for protein structures is important for the study of protein geometry, which, to a large extent, determine the functions of proteins. There are a number of approaches for modelling; one might focus on the conformation of the backbone or H-bonds, and the model may be based on the geometry or the topology of the structure in focus. We focus on the topology of H-bonds in proteins, and explore the link between the topology and the geometry of protein structures. More specifically, we take inspiration from CASP Evaluation of Model Accuracy and investigate the extent to which structural similarities, via GDT\_TS, can be estimated from the topology of H-bonds. We report on two experiments; one where we attempt to mimic the computation of GDT\_TS based solely on the topology of H-bonds, and the other where we perform linear regression where the independent variables are various scores computed from the topology of H-bonds. We achieved an average $\Delta\text{GDT}$ of 6.45 with 54.5\% of predictions inside 2 $\Delta\mathrm{GDT}$ for the first method, and an average $\Delta\mathrm{GDT}$ of 4.41 with 72.7\% of predictions inside 2 $\Delta\mathrm{GDT}$ for the second method.
\end{abstract}
\bigskip
\end{@twocolumnfalse}
]

\section{Introduction}
It is widely recognised that the diverse functions of proteins are highly dependent on the three-dimensional structures of their native conformations. An effective model for describing the geometric structures of proteins is therefore important for the study of protein structures. One of the earliest models for describing the backbone conformation of proteins is the Ramachandran plots, which plots the dihedral angles $(\varphi, \psi)$ before and after each $C^\alpha$ atoms in two-dimensional distributions \cite{ramachandran63}. The method has since been updated and extended to be used in structural validation \cites{lovell03, read11} and a number of other purposes (see, for example, \cite{carugo13acta} for a review). The extensions to the Ramachandran plots include combining two consecutive pairs of conformation angles \cite{levitt76}, and characterising entire proteins by the averages over $\varphi$ and $\psi$ \cite{carugo13amino}. Another approach is to use the coordinates of backbone atoms instead of dihedral angles. Examples of this approach include the notion of curvature and torsion taken from differential geometry \cites{rackovsky78, rackovsky84}, and projection of nearby atoms to a small sphere centred at each $C^\alpha$ atoms \cite{peng14}. While the above methods are all based on the geometry of the backbone, an alternative approach is possible by considering its topology. A number of studies have used ideas from knot theory to study the link between topology and geometry of proteins \cites{levitt83, chen96, rogen03}. Yet another approach is to focus on H-bonds, which is one of the main mechanisms determining and stabilising the native structure of the proteins \cite{bordo94, rose93, pace14}. Studies suggest incorporating H-bond geometry improves the quality of protein structure models \cites{grishaev04, morozov04}. In \cite{penner14}, spatial rotations were introduced as a systematic three-dimensional descriptor of H-bond geometry, and were found to correspond well to the concrete secondary structures and other local structural motifs. The dataset from \cite{penner14} has further been used by Penner in \cites{penner20, penner21} to estimate free energy of coronarirus spike proteins, with a view to identifying specific sites of interest for vaccine development. If we concentrate on the topology of H-bonds, we obtain a graph, with backbone atoms as vertices and the covalent and H-bonds as edges. Such H-bond graphs were used to study the dynamics of membrane proteins \cite{siemers19}, and for structural comparison \cite{rahat09}. In \cite{penner10}, an extension to this structure was used to study protein structures.

In this paper, we investigate the link between H-bond topology of proteins and their geometric structures. We take inspiration from CASP Evaluation of Model Accuracy (EMA) \cite{cozzetto07}, and investigate how well we can estimate the GDT\_TS of the submitted structures using H-bond graphs of the submitted and target structures. GDT\_TS has been criticised, among others, for being dependent on the lengths of proteins and for having somewhat arbitrary distance cutoffs \cites{xu10, garg16}. Nonetheless it is a widely accepted measure used to compare protein structures, and we use it here as an indication of structural similarities. We designed two experiments. In the first experiment, we attempt to follow the algorithm for computing GDT\_TS, but with only the proteins' topological information (from their H-bond graphs) as the input. The second experiment is a linear regression where independent variables are certain similarity scores computed from the protein H-bond graph, and the dependent variable is GDT\_TS. We note here, that our methods are not intended as an attempt for the CASP EMA. Indeed, both methods require the target structure's H-bond graph as part of the input data, which is not available in CASP. They are intended as an investigation into the usefulness of protein topology in comparison of protein structures. However, one could of course imagine combining our methods with an algorithm to predict H-bond graphs from primary sequences to be used in CASP EMA or similar experiments. Indeed, the idea for the investigation originated in a novel approach to the protein folding problem, inspired in part by \cite{penner14}. It is based on a two-stage process, where in the first stage one or more H-bond graphs are predicted from a primary sequence, then in the second stage the geometric structure is predicted from the H-bond graph(s). We have a method to enumerate possible H-bond graphs, as well as a method to predict local geometric structure of proteins from H-bond graphs \cites{andersen21enum, andersen21pre, andersen21topo}. The current study fits in this programme as a ``proof of concept'' for the idea that H-bond topology of proteins is strongly linked to their geometric structures.  Our model is purely based on the topology of protein structures, therefore is less affected by the dynamic nature of the proteins, which is important in their diverse functions \cite{henzler07}. Furthermore, it is independent of alignment, which simplifies its use in potential high throughput applications.

\section{Methods and Results}

\subsection{Dataset}\label{sec:gdt_data}
\begin{figure}[h]
  \centering
  \includegraphics[width=.9\linewidth]{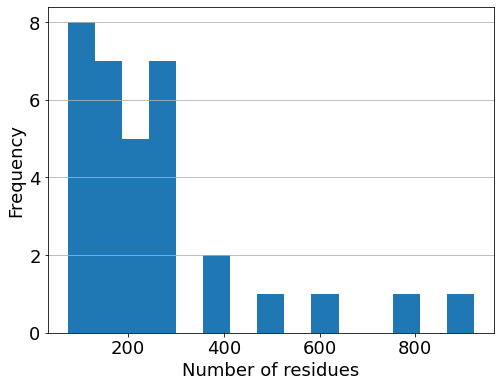}
  \caption[Histogram of target structures]{Frequency of target structures in CASP14 by length.}
  \label{fig:gdt_target14}
\end{figure}

\begin{figure}[h]
  \centering
  \includegraphics[width=.9\linewidth]{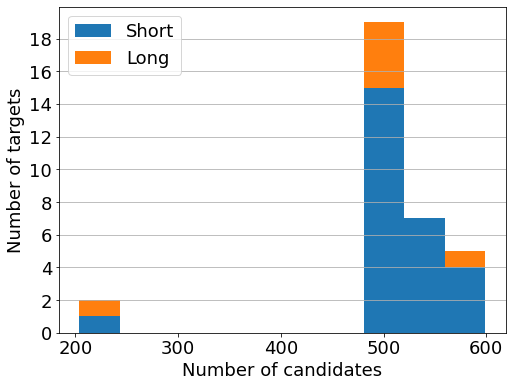}
  \caption[Histogram of candidate structures]{Number of candidates per target structure in CASP14. Short targets are those with fewer than 300 residues, and the long targets are with more than 300 residues.}
  \label{fig:gdt_cand14}
\end{figure}

The dataset consists of 33 target structures for CASP14 together with the submitted candidate structures, downloaded from CASP data archive \cite{casp14data} (There were 34 target structures available for download, but one, T1044, did not have any corresponding candidates and was dropped.). The size of proteins, measured in the number of residues, ranged from 74 to 922 (\Cref{fig:gdt_target14}). Majority of the target structures had length less than 300 residues, with 6 targets having more than 300 residues. The range of the number of candidate structures per protein was from 204 to 599, with the majority of targets receiving more than 500 submissions (\Cref{fig:gdt_cand14}; Participants are allowed to submit more than one candidate structure). The larger target structures seems to have received as many submissions as the smaller target structures. 

\begin{figure}[h]
  \centering
  \includegraphics[width=.9\linewidth]{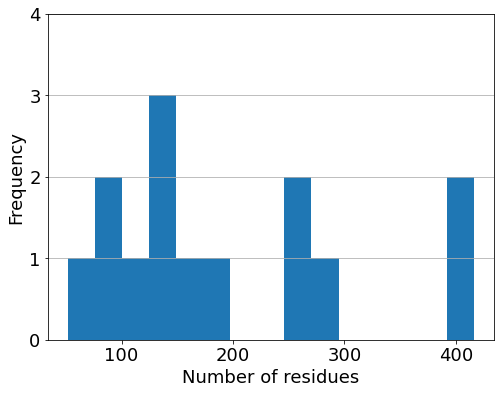}
  \caption[Histogram of target structures]{Frequency of target structures in CASP13 by length.}
  \label{fig:gdt_target13}
\end{figure}

We also utilised data from CASP13 to construct our regression model (\Cref{sec:gdt_linreg}). There were 20 target structures available for download, with length ranging from 52 to 405 residues, and two structures having more than 300 residues (\Cref{fig:gdt_target13})

The data was processed to obtain information about the H-bonds, following the procedure described in \cite{penner14}. The H-bonds were determined by the DSSP program \cite{kabsch83}, with the additional conditions \cite{baker84};
  \begin{align*}
    &\text{HO-distance} < 2.7\text{Å} \\
    &\text{angle(NHO)}, \text{angle(COH)} > 90^\circ.
  \end{align*}

\begin{figure}[h]
    \centering
    \includegraphics[width=.9\linewidth]{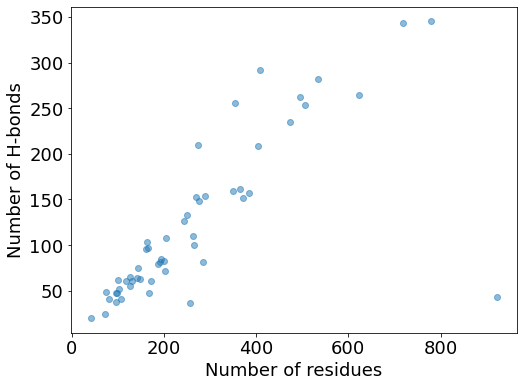}
    \caption[H-bonds vs length]{Number of H-bonds versus length for the target structures in CASP13 \& 14.}
    \label{fig:hbonds_length}
  \end{figure}

  For the majority of proteins in the resulting data, the number of H-bonds was roughly half of the length measured as the number of residues (\Cref{fig:hbonds_length}).

\subsection{GDT-like algorithm based on H-bond graphs}\label{sec:gdt_like}
We attempt to mimic the GDT algorithm \cite{zemla03}, but based only on protein H-bond graphs, i.e.\ based on information about the protein's hydrogen bonds, but not its geometric structure.

Let $T$ be the graph of the target protein, with vertices $\{v_1, \dotsc, v_l\}$ representing the residues, ordered along the backbone, and edges $\{e_1, \dotsc, e_m\}$ representing the backbone peptide bonds and H-bonds. Similarly, let $C$ be the graph of the candidate protein, with vertices $\{w_1,\dotsc,w_l\}$ representing the residues, ordered along the backbone, and edges $\{f_1,\dotsc,f_n\}$ representing primary and H-bonds. For a set $S$, let $\#S$ denote the number of elements in $S$, and for a graph $G$, let $\mathcal{V}(G)$ and $\mathcal{E}(G)$ denote the set of vertices and edges in $G$, respectively. We have $l=\# \mathcal{V} (T)(=\# \mathcal{V}(C))$. The idea is to start with small subgraphs of $C$ and $T$ (corresponding to the same backbone segment), and to ``grow'' them incrementally, until the difference between the subgraphs is over a pre-determined threshold value. We repeat this for different initial subgraphs, and determine the maximum subgraph of $C$, $\hat{C}_{\mathrm{sub}}$, whose difference from the corresponding subgraph of $T$ (the correspondence is defined below in the detailed description) is below some threshold value $r$. The score for the candidate graph, which we call $g_r$, is then given by
\begin{equation}
  \label{eq:gr}
  g_r = 100 \times \frac{\# \mathcal{V}(\hat{C}_{\mathrm{sub}})}{\# \mathcal{V}(C)}.
\end{equation}

We give a detailed description of the algorithm below, together with the pseudocode in \Cref{alg:pcode}. Let $d(A, B)$ be a distance function between graphs $A$ and $B$ and let $r\in (0,\infty)$. We compute the score of similarity between the graphs $T$ and $C$ as follows;

\begin{enumerate}
\item Set $i=1$.
\item \label{forstart} We select a subgraph $C_{\text{sub}}(i)$ of $C$, consisting of three vertices $\{w_i, w_{i+1}, w_{i+2}\}$ starting from the $i$th position and the edges connecting them. Select a subgraph $T_{\text{sub}}(i)$ of $T$ in the same way.
\item Compute the distance measure $d_i= d(T_{\text{sub}}(i), C_{\text{sub}}(i))$.
\item If $d_i \geq r$, where $r$ is the pre-determined limit, the initial segment is already over the limit value. Increment $i$ by 1, go to \ref{forstart}.
\item If $d_i < r$, ``grow'' the subgraph $C_{\text{sub}}(i)$ by adding all edges that are connected to $C_{\text{sub}}(i)$, together with the vertices connected to these edges. Call the selected edges and vertices, together with $C_{\text{sub}}(i)$, $C_{\text{sub}}2(i)$. Select $T_{\text{sub}}2(i)$ from $T$ in the same way.
\item Compute the distance measure $d_i = d(T_{\text{sub}}2(i), C_{\text{sub}}2(i))$.
\item Repeat the above two steps (``growing'' subgraphs and computing the distance measure),  until $d_i \geq r$. If $d_i \geq r$, move to the next starting segments by incrementing $i$ by 1, go to \ref{forstart}.
\item If $C_{\text{sub}}(i) = C$, we have the entire graph under the limit value. 
\item After going through all starting segments, we have a set $S=\left\{C_{\text{sub}}(i) \middle| i \in \{1,\dotsc,l-2\}\right\}$ of maximal $C_{\text{sub}}(i)$'s. Select the longest $C_{\text{sub}}(i)$ in $S$, which we call $\hat{C}_{\text{sub}}$.
\item $g_r = 100 \times \frac{\# \mathcal{V}(\hat{C}_{\text{sub}})}{\# \mathcal{V}(C)}$.
\end{enumerate}

\begin{algorithm}
  \begin{algorithmic}
    \For{$i$ in $\{1,\dotsc,l-2\}$}
    \State Let $C_{\text{sub}}(i)$ be the subgraph of $C$ obtained by taking three vertices $\{w_i, w_{i+1}, w_{i+2}\}$ and the edges connecting them in $C$
    \State Let $T_{\text{sub}}(i)$ be the subgraph of $T$ obtained by taking three vertices $\{v_i, v_{i+1}, v_{i+2}\}$ and the edges connecting them in $T$
    \State Compute the distance measure $d_i = d(T_{\text{sub}}(i), C_{\text{sub}}(i))$
    
    \If {$d_i \geq r$} 
    \State Continue to next $i$
    \EndIf

    \While {True}
    \State Let $T_{\text{sub}}2(i)$ be the subgraph of $T$ obtained by taking $T_{\text{sub}}(i)$ together with all edges connected to the vertices in $T_{\text{sub}}(i)$, and the end-vertices of these edges (i.e.\ "grow" the subgraph by 1 edge+vertex pair)
    \State Let $C_{\text{sub}}2(i)$ be the subgraph of $C$ obtained in the same manner
    \State Compute the distance measure $d_i=d(T_{\text{sub}}2(i), C_{\text{sub}}2(i))$

    \If {$d(i) \geq r$}
    \State Break out of while loop
    \EndIf

    \State Set $T_{\text{sub}}(i) = T_{\text{sub}}2(i)$
    \State Set $C_{\text{sub}}(i) = C_{\text{sub}}2(i)$

    \If {$T_{\text{sub}}(i) == T$} 
    \State Break out of while loop
    \EndIf
    
    \EndWhile
    \EndFor

    \State $\hat{C}_{\text{sub}} = \max\left\{C_{\text{sub}}(i) \middle| i \in \{1,\dotsc,l-2\}\right\}$
    \State $\mathrm{Score} = 100 \cdot \# \mathcal{V}(\hat{C}_{\text{sub}}) / \# \mathcal{V}(C)$
    
  \end{algorithmic}
  \caption{Pseudocode for GDT-like algorithm}
  \label{alg:pcode}
\end{algorithm}

\begin{figure*}[t]
  \centering
  \includegraphics[width=.9\linewidth]{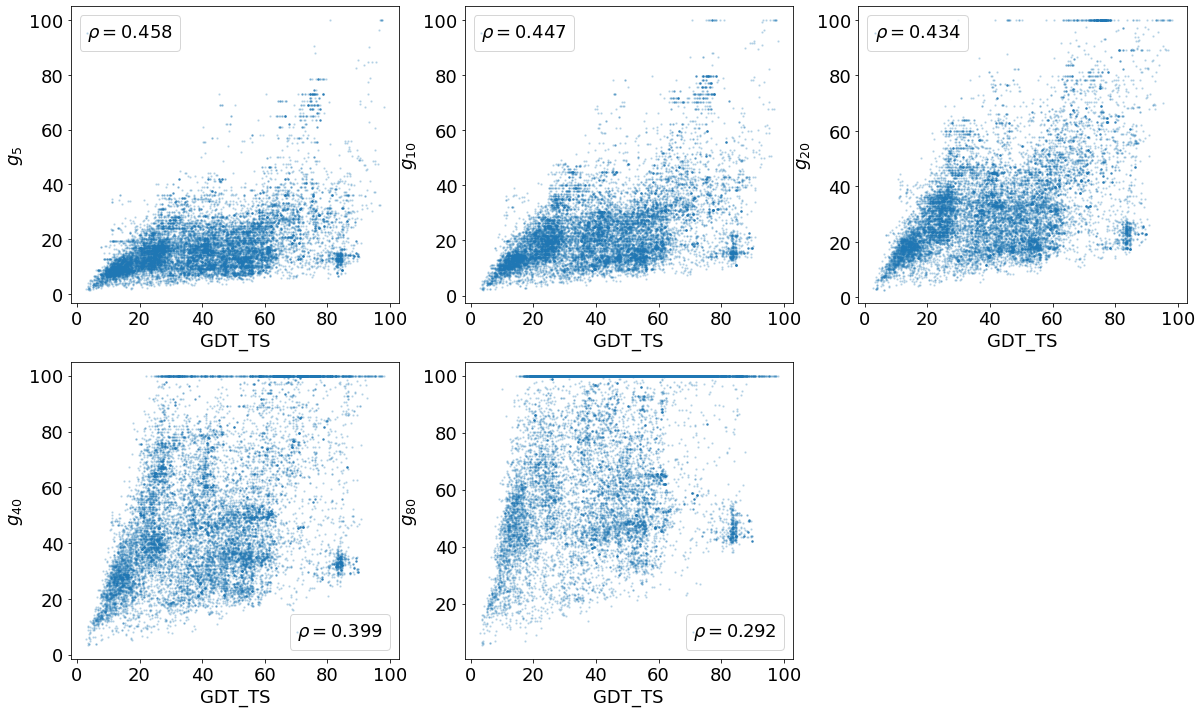}
  \caption[Distribution of $g_r$]{Distribution of $g_r$ scores for different values of $r$, against GDT\_TS. $\rho$ is the Spearman's correlation coefficient between GDT\_TS and $g_r$ scores.}
  \label{fig:gr_dist}
\end{figure*}

\begin{figure*}[t]
  \centering
  \includegraphics[width=.9\linewidth]{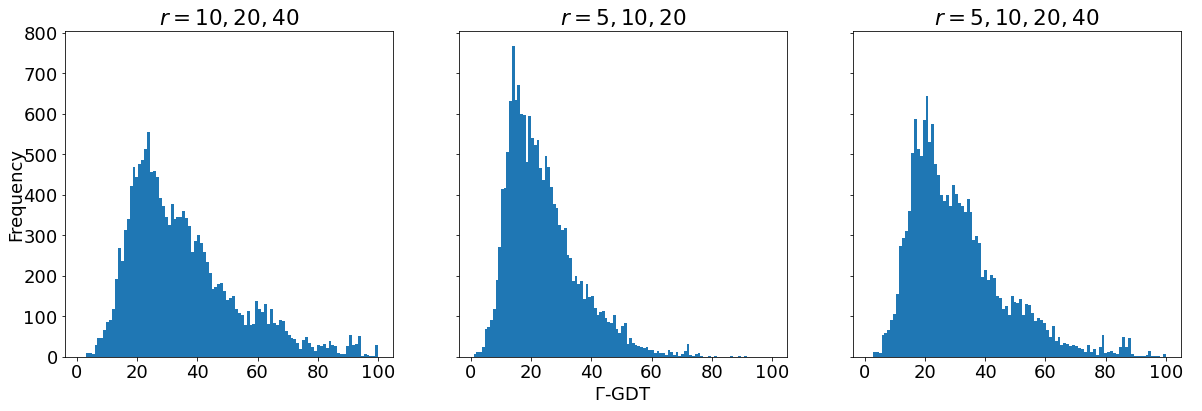}
  \caption[$\Gamma$-GDT distributions]{Distribution of $\Gamma$-GDT for different combinations of $r$-values.}
  \label{fig:ggdt_rs}
\end{figure*}

For the current analysis we define the distance function $d$ by
\begin{equation}
  \label{eq:d}
  d(A,B) = \# (\mathcal{E}(A) \ominus \mathcal{E}(B)),
\end{equation}
where $\ominus$ denotes the symmetric difference of two sets;
\begin{equation*}
  A \ominus B = (A \setminus B) \cup (B \setminus A).
\end{equation*}
The algorithm is also dependent on the limit value $r$ for the distance between two subgraphs. We tested for the effect of different $r$ values by computing the score $g_r$ for $r=5, 10, 20, 40, 80$ for all candidate structures and looking at their distributions, together with their correlation with GDT\_TS (\Cref{fig:gr_dist}). As a result $r=80$ was excluded as being too high (resulting in more than 20\% of all structures having score of 100). GDT\_TS is computed as an average of scores for four different cutoff values (1,2,4, and 8 \AA) \cite{kryshtafovych07}. We imitate this by computing an average over  different sets of $r$-values. The distributions of (average) scores for different sets of $r$-values are shown in \Cref{fig:ggdt_rs}. Based on these, we chose the average over $r$-values 10, 20 and 40 as our final score, since the combination had the widest spread of values. We call the final composite score $\Gamma$-GDT;
\begin{equation}
  \label{eq:g-gdt}
  \mbox{$\Gamma$-GDT} = (g_{10} + g_{20} + g_{40}) / 3.
\end{equation}

\begin{figure}[h]
  \centering
  \includegraphics[width=.9\linewidth]{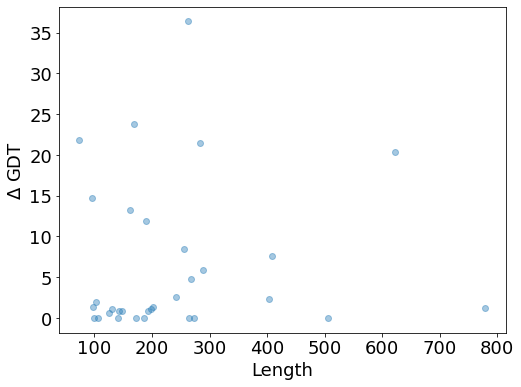}
  \caption[$\Delta\mathrm{GDT-g}$]{Distribution of $\Delta\mathrm{GDT}$ against length (measured in number of residues) for 33 target structures, predicted using $\Gamma$-GDT.}
  \label{fig:dgdt_g}
\end{figure}

\begin{figure}[h]
  \centering
  \includegraphics[width=.9\linewidth]{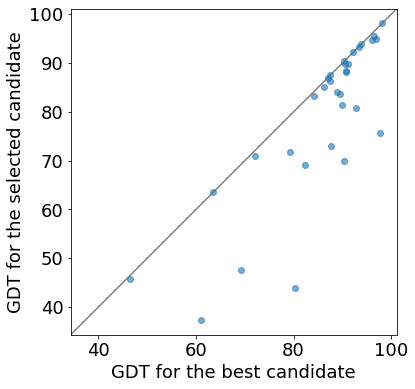}
  \caption[Best GDT vs. selected GDT]{GDT\_TS for the best candidate structure against GDT\_TS for the selected structure.}
  \label{fig:best_selected}
\end{figure}

We then predicted the best candidate structure for each target by selecting the structure with the highest $\Gamma$-GDT, and look at the difference between GDT\_TS of our prediction and GDT\_TS of the best candidate for each target structure, which we call $\Delta\mathrm{GDT}$. The distribution of $\Delta\mathrm{GDT}$ is shown in \Cref{fig:dgdt_g}. The average $\Delta\mathrm{GDT}$ for all targets was 6.45, with the highest value of 36.38. We were able to identify a candidate with $\Delta\mathrm{GDT}<2$ in 18 targets, and with $\Delta\mathrm{GDT}<10$ in 24 targets (\Cref{fig:dgdt_g}). The distribution of GDT\_TS for the best candidate structure against GDT\_TS for the selected structure is shown in \Cref{fig:best_selected}. It turned out the particular set of $r$-values we chose for the computation of $\Gamma$-GDT gives the best prediction result (\Cref{tab:dgdt_rs}).

\begin{table}[h]
  \centering
  \begin{tabular}{l|rr}
    $r$-values & $\Delta\mathrm{GDT}<2$ & $\Delta\mathrm{GDT}<10$ \\
    \hline
    10,20,40 & 18 & 24 \\
    5,10,20 & 14 & 22 \\
    5,10,20,40 & 17 & 23
  \end{tabular}
  \caption[$\Delta\mathrm{GDT}$ and $r$-values]{Number of predictions (out of 33) with specified $\Delta\mathrm{GDT}$ range for different combinations of $r$-values.}
  \label{tab:dgdt_rs}
\end{table}

We also tested for two different distance function to \eqref{eq:d}. In the first, we reduced the contribution made by an edge in the set $\mathcal{E}(C) \ominus \mathcal{E}(T)$, if there is an edge that lies close to it. For $x \in \R$, define a function $f_1$ by
\begin{equation}
  \label{eq:f1}
  f_1(x) =
  \begin{cases}
    1 & \text{if } \abs{x} > 4 \\
    \abs{x}/4 & \text{otherwise}.
  \end{cases}
\end{equation}
We define a new distance function $d_1$ by
\begin{align}
  \label{eq:d1}
  d_1&(A,B) =\nonumber\\
           & \sum_{(p,q) \in A \setminus B} \min \{\frac{f_1(p-p')+f_1(q-q')}{2}\nonumber\\
     &\qquad\qquad | (p',q')\in B \setminus A\} \nonumber\\
           & + \sum_{(p,q) \in B \setminus A} \min \{\frac{f_1(p-p')+f_1(q-q')}{2}\nonumber\\
           &\qquad\qquad| (p',q') \in A \setminus B\}.
\end{align}
In the second, we tried to reduce the contribution by a ``close'' edge further by setting
\begin{equation}
  \label{eq:f2}
  f_2(x) = \begin{cases}
    1 & \text{if } \abs{x} > 4 \\
    \frac{\exp(\abs{x})-1}{\exp(4)-1} & \text{otherwise},
  \end{cases}
\end{equation}
and
\begin{align}
  \label{eq:d_2}
  d_2&(A,B) =\nonumber\\
           & \sum_{(p,q) \in A \setminus B} \min \{\frac{f_2(p-p')+f_2(q-q')}{2}\nonumber\\
           &\qquad\qquad | (p',q')\in B \setminus A\} \nonumber\\
           & + \sum_{(p,q) \in B \setminus A} \min \{\frac{f_2(p-p')+f_2(q-q')}{2}\nonumber\\
           &\qquad\qquad| (p',q') \in A \setminus B\}.
\end{align}

\begin{figure}[h]
      \centering
      \includegraphics[width=.9\linewidth]{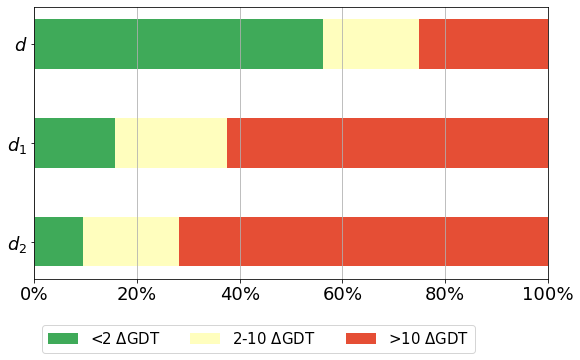}
      \caption[dgdt and distance functions]{Percentages of predictions with $\Delta\mathrm{GDT}<2$, $2 \leq \Delta\mathrm{GDT} < 10$, and $\Delta\mathrm{GDT} \geq 10$ for different distance functions.}
      \label{fig:dgdt_ds}
    \end{figure}

The prediction results for different distance functions are shown in \Cref{fig:dgdt_ds}. We see that the performance of the original distance function $d$ \eqref{eq:d}, which is a simple count of the elements in the symmetric difference of the sets of edges, is significantly better than the two modified distance functions.

\subsection{Linear regression based on the protein fatgraph model}\label{sec:gdt_linreg}
The second method is a linear regression on the similarity scores which we compute based on the hydrogen bonds in the candidate and target structures. Each hydrogen bond is identified by the position of its donor- and acceptor atoms, so each bond can be expressed as a 2-tuple of integers $(p,q)$, where the donor is the $p$'th atom along the backbone and the acceptor the $q$'th. 

The first of our similarity scores is the proportion of the bonds, which are correctly identified in the candidate structure. In other words, if $H_T, H_C$ are the sets of hydrogen bonds respectively in the target structure and in the candidate structure, then the first score $P$ is defined as;
\begin{equation*}
  P = \frac{\# (H_T \cap H_C)}{\# H_T},
\end{equation*}
where we use the fact that for two bonds $(p,q)$ and $(p',q')$, $(p,q)=(p',q')$ iff $p=p'$ and $q=q'$.

The second similarity score $S_n$ depends on a parameter $n \in \mathbb{N}$. For a non-negative integer $x \in \mathbb{Z}$, define
\begin{equation}
  \label{eq:score}
  f_n(x) =
  \begin{cases}
    1 - x/n & \text{ if } x \leq 2n \\
    -1 & \text{ otherwise}
  \end{cases}.
\end{equation}
For a bond $(p,q) \in H_C$, set
\begin{align*}
  s_C&((p,q)) \\
     &= \max \left\{f_n(\abs{p-p'}) + f_n(\abs{q-q'}) \right. \\
     & \qquad\qquad\qquad\qquad \left.\mid (p',q') \in H_T \setminus H_C \right\}.
\end{align*}
Similarly for $(p,q) \in H_T$, set
\begin{align*}
  s_T&((p,q)) \\
     &= \max \left\{f_n(\abs{p-p'}) + f_n(\abs{q-q'}) \right. \\
     & \qquad\qquad\qquad\qquad \left.\mid (p',q') \in H_C \setminus H_T \right\}.
\end{align*}
$S_n$ is then given by
\begin{align*}
  S_n &= \frac{1}{\# ((H_T \setminus H_C) \cup (H_C \setminus H_T))} \times \\
      &\left(\sum_{(p,q) \in H_C \setminus H_T}s_C((p,q)) + \right.\\
      &\left.\quad\sum_{(p',q') \in H_T \setminus H_C}s_T((p',q')) \right).
\end{align*}
So for a given candidate structure, we can compute $S_n$ for different $n$'s.

\begin{figure}[h]
  \centering
  \includegraphics[width=\linewidth]{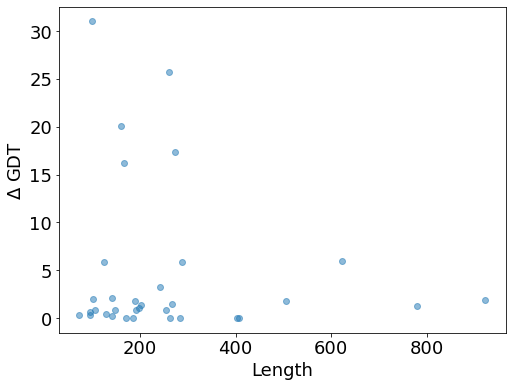}
  \caption[$\Delta\mathrm{GDT}$]{Distribution of $\Delta\mathrm{GDT}$ against length (measured in number of residues) for 33 target structures, by linear regression method.}
  \label{fig:dgdt}
\end{figure}

Having calculated $P$ and $S_n$, $n \in I$, where $I$ is a subset of $\{1,2,\dotsc,10\}$, for all candidate structures, we perform a linear regression with $P$, $S_n$ as independent variables and GDT\_TS as the dependent variable. We estimated the regression model using data from CASP13, and applied the model to data from CASP14. After testing for all subsets $I \subset \{1,2,\dotsc,10\}$ by running multiple regression with $P$ and $S_n, n \in I$ as independent variables, we found that setting $I = \{ 2 \}$ gave the best results with CASP13 data. The regression equation, based on all candidate structures in CASP13, was determined to be
\begin{equation*}
  \mathrm{GDT\_TS} = 10.70 + 0.63P + 1.26S_2.
\end{equation*}
Using this equation, we estimated the GDT\_TS for CASP14 candidate structures, and selected the structure with the highest estimated GDT\_TS for each target. We were able to identify a candidate structure with $\Delta\mathrm{GDT} < 2$ for 23 out of 33 targets, with the average $\Delta\mathrm{GDT}$ of 4.57 (\Cref{tab:reg_result}). The frequency distribution of $\Delta\mathrm{GDT}$ is shown in \Cref{fig:dgdt}. The large $\Delta\mathrm{GDT}$ values were observed in shorter proteins, although it must be noted that most of targets have lengths less than 300 residues. 

We also investigated the effect of the score function \eqref{eq:score} by scaling it with the exponential function;
\begin{equation}
  \label{eq:score2}
  \tilde{f}_n(x) =
  \begin{cases}
    1 - \frac{2(\exp(x)-1)}{\exp(2n)-1} & \text{ if } x \leq 2n \\
    -1 & \text{ otherwise}
  \end{cases}.
\end{equation}
Compared to \eqref{eq:score}, the new score function \eqref{eq:score2} gives smaller penalties to difference in bond positions, especially when the difference is small. Using the data from CASP13 and \eqref{eq:score2}, we found that setting $I = \{2,6,8,10\}$ gave the best result with regards to identifying the most candidates with $\Delta\mathrm{GDT}<2$. However, the $S_n$ scores are strongly correlated, and we decided to use $I = \{2\}$ again. This gave a result close to that obtained with $I = \{2,6,8,10\}$ (11 targets with $\Delta\mathrm{GDT}<2$, compared to 12 targets with $\Delta\mathrm{GDT}<2$). The regression equation was
\begin{equation*}
  \mathrm{GDT\_TS} = 9.68 + 0.63P + 0.20S_2.
\end{equation*}
The new score function \eqref{eq:score2} resulted in a small improvement for prediction with CASP14 data, where we were able to identify a candidate structure with $\Delta\mathrm{GDT}<2$ for 24 out of 33 targets, and the average $\Delta\mathrm{GDT}$ of 4.41 (\Cref{tab:reg_result}).

We then removed the $S_n$ scores from the independent variables, and ran the regression with only $P$ scores as the independent variable. The regression equation now read
\begin{equation*}
  \mathrm{GDT\_TS} = 9.56 + 0.63P.
\end{equation*}
This only resulted in a small drop in our ability to identify the best candidate structure, with $\Delta\mathrm{GDT}<2$ for 22 out of 33 targets and the average $\Delta\mathrm{GDT}$ of 5.53 (\Cref{tab:reg_result}).

\begin{table}[h]
  \centering
  \begin{tabular}{ccc}
    \hline
    & \multicolumn{2}{c}{\% of candidates with} \\
          & $\Delta\mathrm{GDT}<2$ & $\Delta\mathrm{GDT}<10$ \\
    \hline
    $f_n$ & 69.70 & 84.85 \\
    $\tilde{f}_n$ & 72.73 & 84.85 \\
    No $S_n$ & 66.67 & 84.85 \\
    \hline
  \end{tabular}
  \caption[Regression results]{Prediction results for different score functions, showing the percentages of targets (out of 33), where the selected candidate structure had $\Delta\mathrm{GDT}$ less than 2 and 10, respectively.}
  \label{tab:reg_result}
\end{table}

\section{Discussion}\label{sec:gdt_disc}
\begin{figure}[h]
  \centering
  \includegraphics[width=\linewidth]{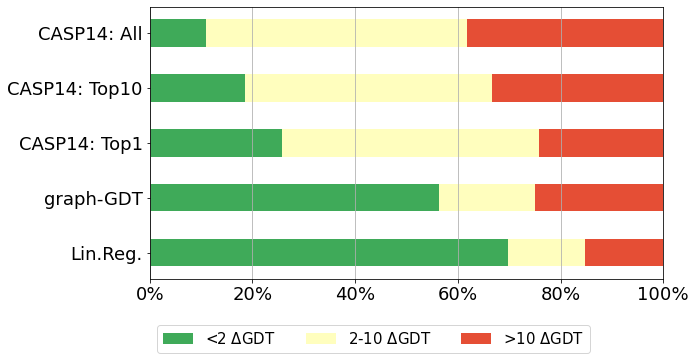}
  \caption[$\Delta\mathrm{GDT}$ bins]{Percentages of the predictions with $\Delta\mathrm{GDT}<2$, $2 \leq \Delta\mathrm{GDT} < 10$ and $\Delta\mathrm{GDT} \geq 10$. The figures for CASP14 are averages of all models, top 10 models, and the best-performing model, ordered by the number of predictions with $\Delta\mathrm{GDT}<2$. The data for CASP14 was obtained from CASP Data Archive (\url{https://www.predictioncenter.org/download_area/CASP14/}) and processed by the authors.}
  \label{fig:dgdt_bins}
\end{figure}

We have shown that the information on H-bonds alone can, to a large extent, correctly assess similarities in geometric structures of proteins. Even though a direct comparison between our results and CASP EMA is not possible, the performance of our methods, measured as the percentages of predictions with $\Delta\mathrm{GDT}<2$ and $\Delta\mathrm{GDT}<10$, are clearly numerically superior to the best performance in CASP 14 EMA (\Cref{fig:dgdt_bins}). It could be argued that both our methods essentially rely on simply counting the matched (or unmatched, in the case of $\Gamma$-GDT) H-bonds in two structures. The modified distance functions in the $\Gamma$-GDT and the $S$ scores in the linear regression, which measures the differences between unmatched H-bonds, have negative or relatively small positive effect on the overall accuracy of predictions. The fact that these relatively simple methods can nonetheless assess similarities in protein structures correctly, demonstrates the strong link between the topology and the geometry of proteins.

In the linear regression analysis, we observe that the larger values of $n$ in $S_n$ scores, which in effect enlarges the search window for ``similar'' H-bonds, do not improve the prediction accuracy.  An explanation could be that it is simply a consequence of using GDT\_TS as the measure of structural similarity, as a small local difference (e.g. an extra turn where there should be none) can result in a significantly lower GDT\_TS. Further investigation is needed to ascertain the cause of this behaviour. 

There are broadly two types of methods used in CASP model accuracy estimation. Consensus, or clustering methods take multiple candidate structures as input and tries to identify a structure, that is the ``best match'' for the input structures according to some criteria. Single-model methods, on the other hand, takes a single candidate structure as an input and tries to estimate its accuracy, independent of other candidate structures. The consensus methods have generally outperformed the single-model methods, and this resulted in the development effort being concentrated on the consensus methods in the past \cite{ray12}. More recently the single-model methods have received more attention and development effort \cites{kryshtafovych18}, as the potential issues with the consensus methods are recognised. One issue, for example, is that the consensus methods may not be very useful in the environment outside the CASP-setup, where a large number of candidate structures may not be available for the input. Another potential issue, related to the first, is that the consensus methods may simply be taking advantage of the fact that many CASP models are now able to produce  high-quality candidate structures, which are, naturally, similar to each other \cite{won19}. Our method is, by construction, unlikely to be improved to outperform the best accuracy estimation methods, as it ignores the geometric data in the candidate structures and only utilises the topological data. However, the relative simplicity of our method means it should be relatively easy to combine it with an existing method to improve its performance. We chose not to attempt it in this paper, as our focus here has been to investigate the link between the topology and the geometry of proteins, rather than to participate in CASP EMA. Nonetheless, as we mentioned in Introduction, one could easily imagine combining our method with an algorithm for predicting hydrogen bonds from a primary sequence. When a high-accuracy prediction of hydrogen bonds becomes possible, our method has the advantage that it could be combined with both a consensus method and a single-model method.

\subsection*{Acknowledgement}
This paper is partly a result of the ERC-SyG project, Recursive and Exact New Quantum Theory (ReNewQuantum) which received funding from the European Research Council (ERC) under the European Union's Horizon 2020 research and innovation programme under grant agreement No. 810573.

\printbibliography

\end{document}